\documentclass[final,number,sort&compress,5p,times,twocolumn]{elsarticle}

\usepackage{graphicx}
\usepackage{amssymb}
\usepackage{amsmath}
\usepackage{lineno}

\usepackage{subfig}
\usepackage{multirow}
\usepackage{sidecap}

\usepackage[colorlinks=true, pdfstartview=FitV, linkcolor=blue,citecolor=blue, urlcolor=blue]{hyperref}
\usepackage{url}
\usepackage{tabularx}
\usepackage{color}
\usepackage{pifont}

\journal{Nuclear Instruments and Methods A}

\begin{document}

\begin{frontmatter}

\title{A measurement of the energy and timing resolution of GlueX Forward
  Calorimeter using an electron beam}

\author[label_IU]{K.~Moriya}
\author[label_IU]{J.P.~Leckey}
\author[label_IU]{M.R.~Shepherd}
\author[label_IU]{K.~Bauer}
\author[label_IU]{D.~Bennett}
\author[label_IU]{J.~Frye}
\author[label_JLab]{J.~Gonzalez}
\author[label_IU]{S.~J.~Henderson}
\author[label_JLab]{D.~Lawrence}
\author[label_IU]{R.~Mitchell}
\author[label_JLab]{E.~S.~Smith}
\author[label_IU]{P.~Smith}
\author[label_JLab]{A.~Somov}
\author[label_JLab]{H.~Egiyan}

\address[label_IU]{Indiana University, Bloomington, IN 47405, USA}
\address[label_JLab]{Thomas Jefferson National Accelerator Facility, Newport News, VA 23606, USA}

\begin{abstract}
The performance of the GlueX Forward Calorimeter was studied using a small version of
the detector and a variable energy electron beam derived from the Hall B tagger at Jefferson Lab.
For electron energies from 110 MeV to 260 MeV, which are near the lower-limits of the design 
sensitivity, the fractional energy resolution was measured to range from $20\%$ to $14\%$,
which meets the design goals.
The use of custom 250~MHz flash ADCs for readout allowed precise
measurements of signal arrival times. The detector achieved timing
resolutions of $0.38$ ns for a
single $100$~mV pulse, which will allow timing discrimination of photon beam bunches
and out-of-time background during the operation of the GlueX detector.
\end{abstract}


\end{frontmatter}

\section{Introduction}
\label{section:Introduction}

The upcoming GlueX Experiment~\cite{GlueX-URL} in Hall D at the Thomas
Jefferson National Accelerator Facility (JLab) is an
experiment that will primarily search for mesons with exotic quantum
numbers.  Meson states of interest will be produced using a photon beam
incident on a proton target.  It is crucial for the experiment to have a detector with good
resolution and high acceptance for multi-particle events --
studying the angular distributions of decay products is an essential tool
in identifying underlying structure of the produced mesons.
The Forward Calorimeter (FCAL) will
be an essential detector for the experiment, providing energy
measurements and timing information for photon showers in the forward
region with polar angles $\theta < 12^{\circ}$ and energies between
100~MeV and 5~GeV.

In the fall of $2011$, a
25-element miniature version of the FCAL was constructed for a beam test
underneath the existing photon tagger of Hall B~\cite{CLAS-URL} at
JLab. The main goals were to verify, as expected, that
the  hardware configuration of the FCAL modules would 
meet the desired energy and timing resolutions.  While many studies with
prototypes were performed during the design phase, this represented
the first test of the production hardware and data acquisition system 
in a beam.  Below we discuss the results of this beam test and compare 
them with previous measurements.

\section{Setup of Experiment}
\label{section:Setup}

\subsection{FCAL components}

\begin{figure}[h!t!b!p!]
  \centering
  \includegraphics[width=0.95\linewidth]{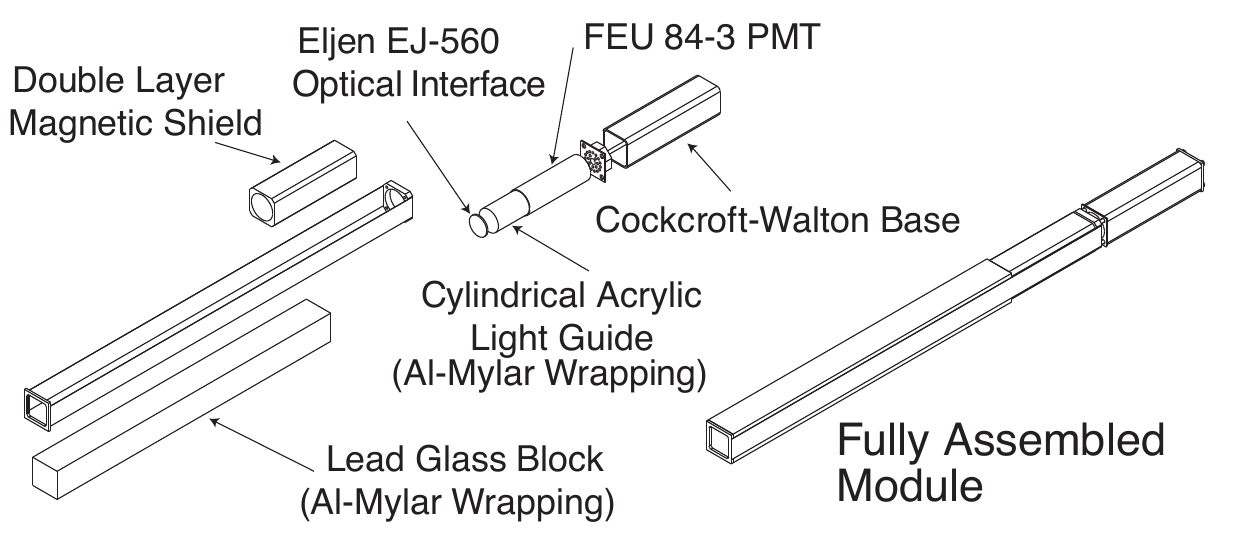}
  \caption{Expanded view of a single module of the GlueX FCAL.}
\label{fig:FCAL_exploded}
\end{figure}

The FCAL for the GlueX experiment consists of $2{,}800$ lead glass
modules, each coupled to its own type FEU 84-3 photomultiplier tube
(PMT) and Cockcroft-Walton base (similar to that detailed in Ref.~\cite{Brunner}). 
The lead glass blocks were equivalent to type F8
manufactured by the Lytkarino Optical Glass Factory~\cite{Lytkarino},
and each have transverse dimensions of $4 \times 4$ $\mathrm{cm}^{2}$
and are $45$ cm long. The Cherenkov light emitted
by the electromagnetic showers produced within the lead glass blocks
will be detected by the PMTs.  The resulting PMT current pulses are
digitized by 12-bit $250$ MHz flash analog-to-digital converters
(fADCs) designed by JLab~\cite{fADC}. Figure~\ref{fig:FCAL_exploded} shows an expanded
view of one of the $2{,}800$ FCAL modules.

The lead glass blocks and most of the PMTs are the same
as those used in previous experiments, E852 at Brookhaven National
Laboratory~\cite{Brabson, Crittenden} and the RadPhi Experiment at
JLab~\cite{Jones_NIM, Jones_NIM2}.

In the GlueX experiment, magnetic shielding of the PMTs is necessary due to the
stray field of the solenoid magnet of up to $200$~G. The PMTs are
placed well within a double layer of soft iron and mu-metal
shield. The light collection from the lead
glass block to the PMT is facilitated using a cylindrical acrylic light guide
glued to the PMT.  An Eljen EJ-560 optical interface ``cookie" is used to connect the
guide to the lead glass block. The primary goal
for this beam test is to verify that we can achieve the design energy
resolution at low energies, where statistical fluctuations in the number
of Cherenkov photons detected dominate the resolution.  This, in turn,
would validate the design of the light collection system using optical
photons with the characteristic Cherenkov angular distribution inside of the
block, a task that had only been performed with ray-tracing simulation.

Besides measuring the energy of electromagnetic showers, the FCAL will
measure the timing of the showers. During GlueX running, the JLab
Continuous Electron Beam Accelerator Facility (CEBAF) beam will
provide an electron beam bunch every $2$~ns. Electrons in the bunch
interact with a thin diamond crystal radiator to produce a linearly-polarized
photon beam.  During nominal running configuration the diamond is oriented
so that the coherent bremsstrahlung process produces a peak in both linear 
polarization and photon flux at photon energies
of about 9~GeV. The photon energy is measured by measuring the 
momentum of the recoil electron using the Hall D tagger. The
photon rates are expected to be up to $10^{8}$ $\text{s}^{-1}$ in the
coherent peak between $8.4$ and $9.0$~GeV, leading to a
high rate of signal photons as well as electromagnetic background. The
timing measurements of the FCAL will allow identification 
of the respective beam bunch that created the photon of interest and also help
in reducing accidental backgrounds.

\subsection{Setup of Detector}

\begin{figure}[h!t!b!p!]
  \centering
  \includegraphics[width=0.95\linewidth]{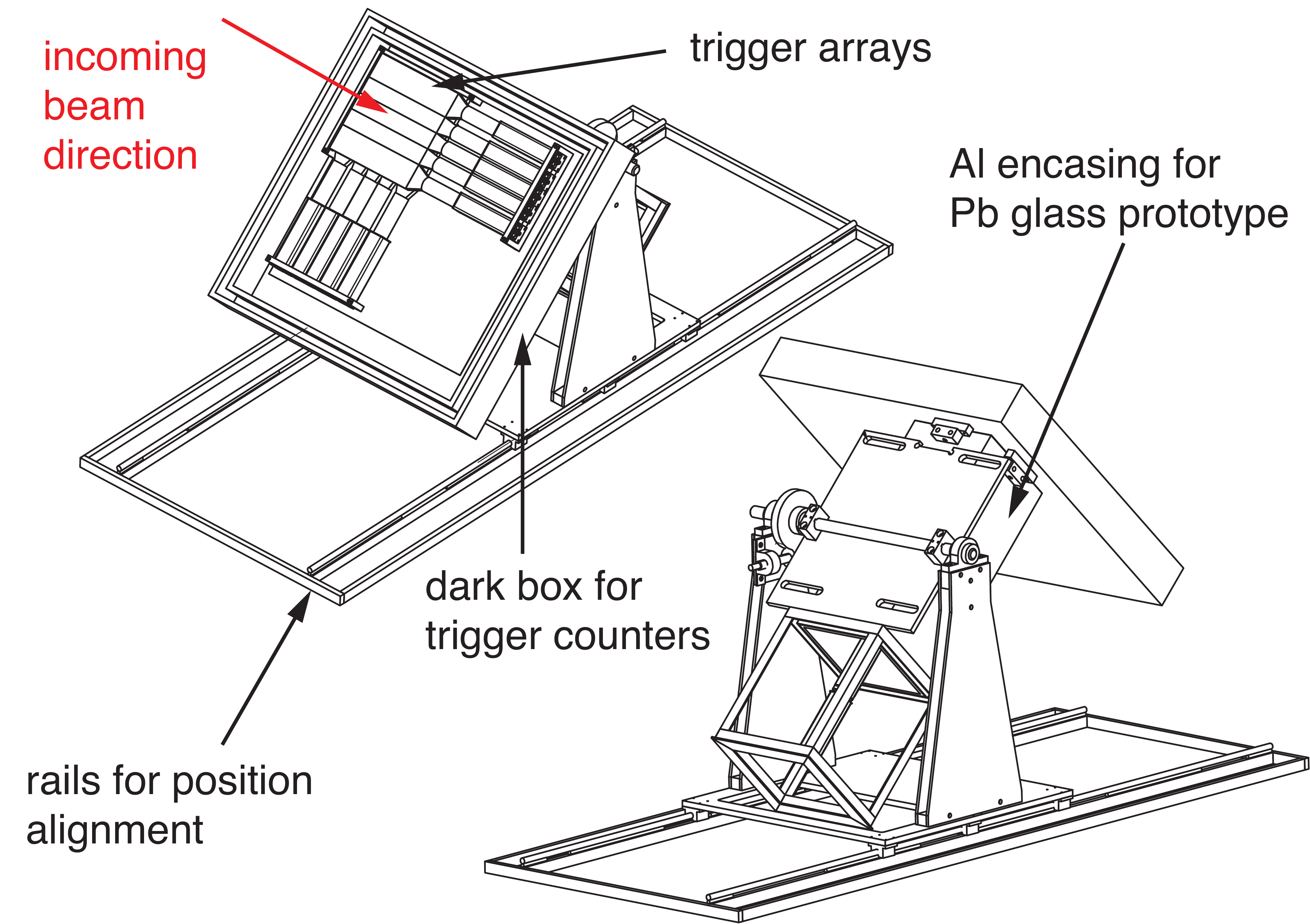}
  \caption{Drawing of the array from the front and back with
    the trigger box attached. For illustration purposes, the lid of the
    trigger box is shown open.}
\label{fig:detector}
\end{figure}

A $5 \times 5$ array of FCAL modules was constructed at Indiana University. 
The modules were
encased in a light-tight aluminum body, with a gear drive
that allowed the modules to be tilted at an angle so that the front
face of the modules were perpendicular to the incoming electron
trajectories. Figure~\ref{fig:detector} shows a drawing of the
detector.

\begin{figure}[h!t!b!p!]
  \centering
  \includegraphics[width=0.95\linewidth]{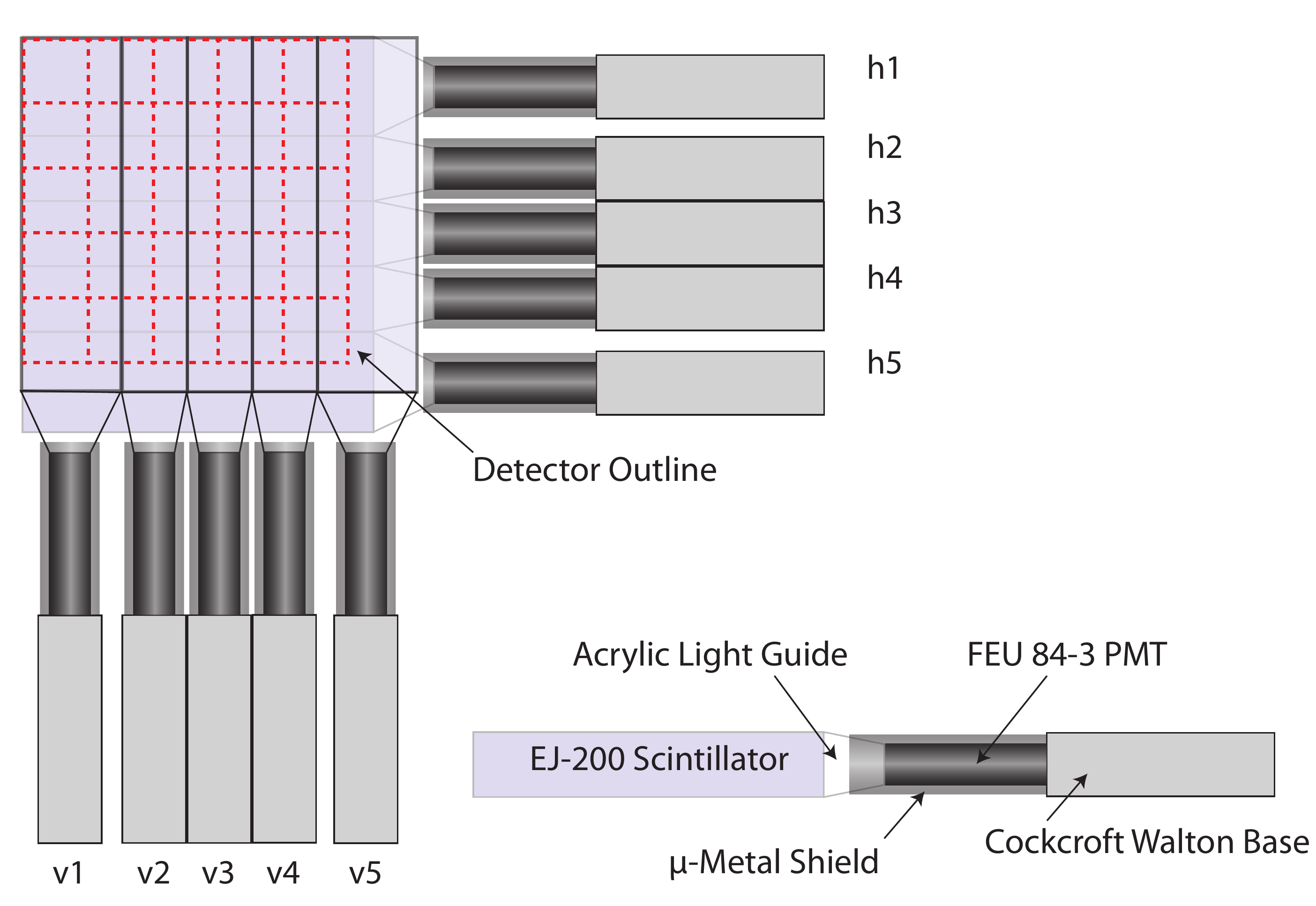}
  \caption{The placement of the trigger scintillator paddles on top of
    the modules. The position of the front face of the $5 \times 5$
    array of detector modules is shown as the dashed lines.}
\label{fig:trigger_positions}
\end{figure}

For triggering purposes, two arrays of five Eljen EJ-200 plastic
scintillator paddles were placed at the front of the detector. The
three inside paddles had widths of $4$ cm, while the two outside paddles 
were $6$ cm wide and used as veto signals to
the trigger. The two arrays were 
oriented perpendicularly to each other, with the centers of each paddle
centered on the border of two modules, as shown in
Fig.~\ref{fig:trigger_positions}.  The scintillators spanned the length
of the perpendicular
array of modules. When looking into the detector
from the upstream end (the ``beam's eye view"), the horizontal paddles were
labeled h1--h5 from the top, and the vertical paddles were labeled
v1--v5 from the left. Each of the trigger paddles were optically
connected to the same type FEU 84-3 PMTs used for reading out the
detector modules. The signals from the PMTs were converted into
NIM signals by a discriminator,
and the digital output of the discriminators were recorded into the
data stream using the same fADCs that were used for recording the
pulse height of the lead glass modules. The trigger arrays were placed
in a dark box that was attached to the 
front of the main detector. The dark box was made of $1.27$ cm-thick
plywood and painted black, with a thin ($1.27$ mm) aluminum
plate underneath the trigger setup to hold the trigger paddles
in place.

In addition to the main trigger counters, there was a thin remote paddle located
several meters upstream of the detector, along the nominal
trajectory of electrons hitting the detector (shown in
Fig.~\ref{fig:HallB}). The remote paddle was
made of a $0.125$ cm-thick plastic scintillator, wrapped in aluminized
mylar, and connected to a Photonis XP2020 PMT. The signal
was also converted into a NIM pulse signal using a discriminator and read
out with a fADC channel. The purpose of this remote paddle was to
identify electrons following the nominal electron trajectory during off-line
analysis.

\begin{figure}[h!t!b!]
\centering
    \includegraphics[width=\linewidth]{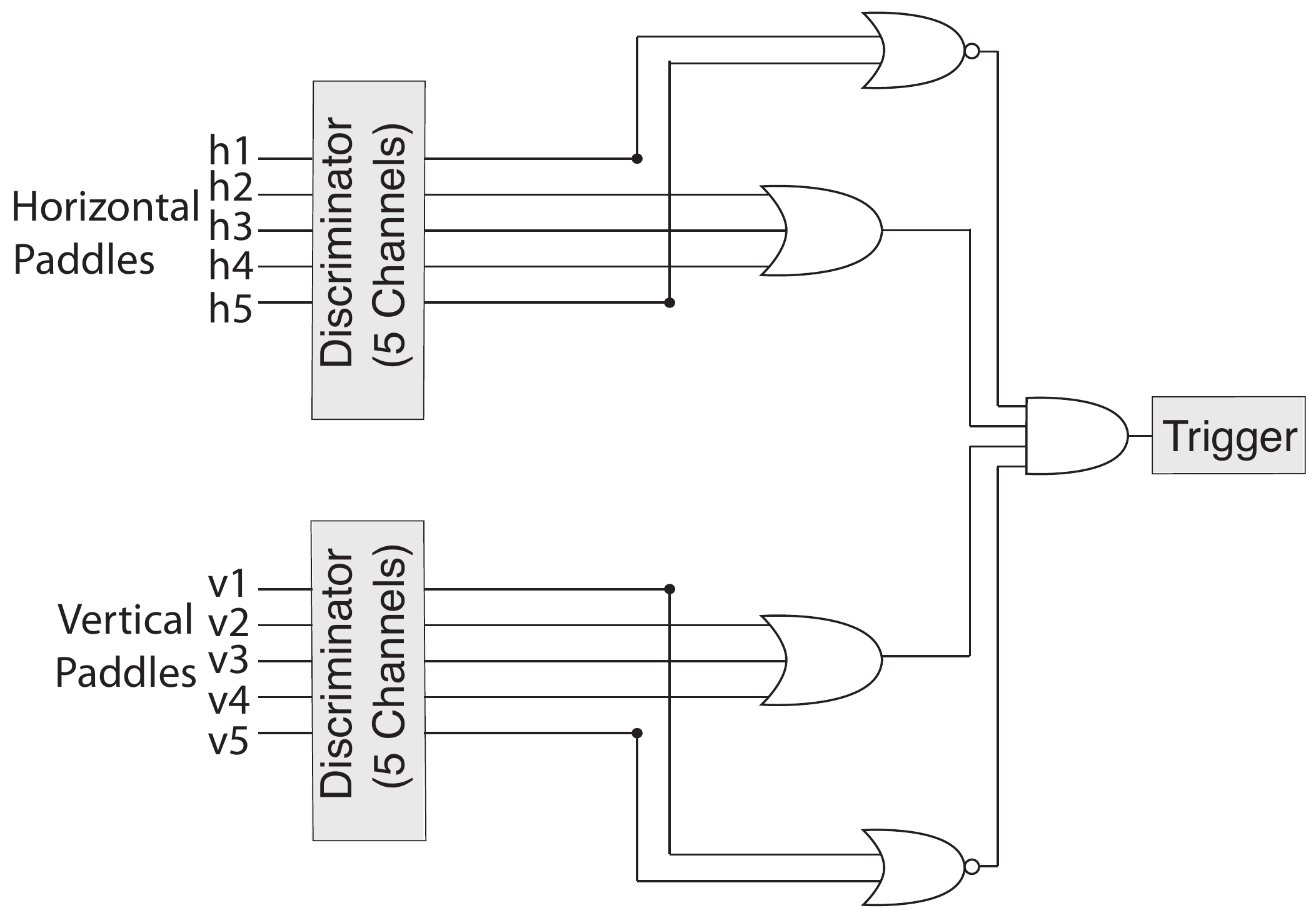}
    \caption{Electronics diagram for the trigger.}
\label{fig:trigger_electronics}
\end{figure}

Figure~\ref{fig:trigger_electronics} shows a diagram of our
trigger logic, which was handled by standard NIM modules. The PMT pulses
from each of the $25$ modules within the detector array were digitized 
using a fADC. To trigger the readout of an event
event to disk, we required a hit within at least one of the inside paddles
from each array. The outside paddles were used as vetoes, and the veto
signal was also individually recorded.

\subsection{Test Setup in Hall B}
\label{subsection:Setup:HallB}

\begin{figure}[h!t!b!p]
  \centering
  \includegraphics[width=0.95\linewidth]{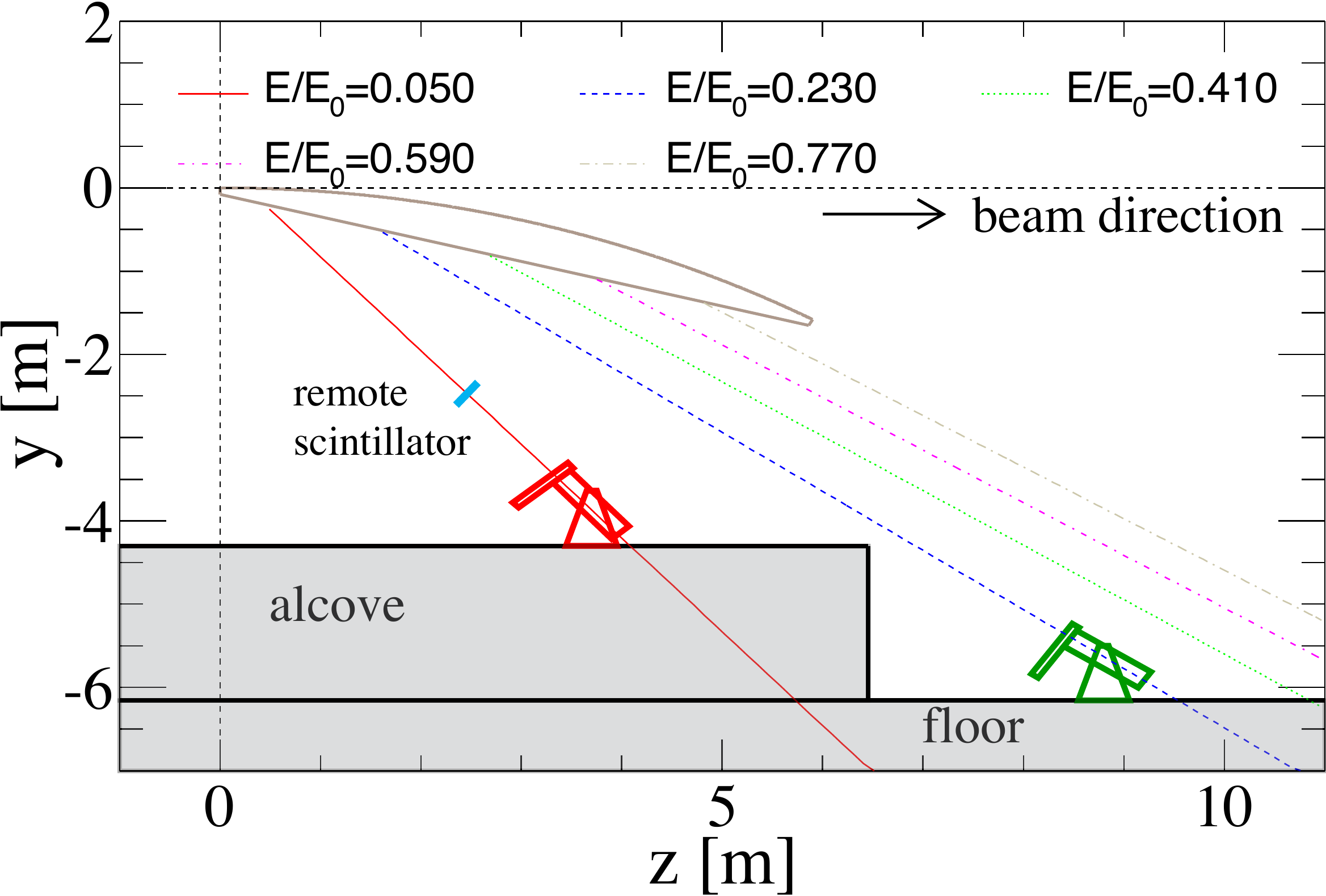}
  \caption{Schematic of the Hall B and the CLAS tagger system. The
    electron beam radiates a photon at coordinates $(0,0)$.
    The trajectories of degraded electrons with varying energies are given by the
    lines originating from the vacuum exit window.
    Depictions of our
    detector are shown in the alcove area and floor.}
\label{fig:HallB}
\end{figure}

The beam test was conducted within Hall~B at JLab, which housed the
CLAS detector and its photon tagger setup. During our beam test, the
CLAS Collaboration was running the HD-ICE experiment with a real photon beam
produced by bremsstrahlung, allowing us to use the electrons that
had radiated photons. Our beam test was conducted parasitically to this
run, and therefore we had no control over the beam energy or intensity.
Figure~\ref{fig:HallB} shows
a schematic of Hall B and the tagger setup. The electron beam from the
accelerator comes in from the left and is incident on a very thin
radiator foil. Immediately following the radiator is the CLAS tagger
magnet, which is a uniform-field dipole magnet.
The electron trajectories are determined by the
energy of the electron, and the position and angle at the
exit of the magnet can be calculated. Details of the CLAS
tagger setup can be found in Ref.~\cite{tagger}.

The electrons exit the vacuum of the tagger magnet through a very
thin exit window~\cite{vacuum_window} and pass through the CLAS tagger
detectors called the E-counters and T-counters. The E-counters are a plane of
$384$, $4$~mm-thick plastic scintillators that detect the position of
each electron for energy identification. The T-counters form a second
plane of $2$~cm-thick plastic scintillators that are used for timing
identification of the electrons. The two detector planes are separated
by $20$~cm, and each counter within each plane is rotated so that it
is facing normal to the incoming trajectory of electrons.

In our beam test, the detector was placed mainly in the
area seen in the left side of Fig.~\ref{fig:HallB} that we call
the alcove. The detector was placed at the calculated location that
intercepted electrons with $5\%$ of the accelerator beam energy and was
angled to face the incoming electrons. We also placed the detector in
the floor area (right side of Fig.~\ref{fig:HallB}), where data were
collected at a higher electron energy fraction. Due to mechanical interference
and running constraints, we could not properly align the detector in
the floor area, and hence did not use those runs for our energy resolution
measurements (Sec.~\ref{section:EnergyResolution}). However, since
the timing resolution studies (Sec.~\ref{section:TimingResolution})
are not very sensitive to the detector alignment, we utilized these runs for
this purpose, as they had the largest range of signal sizes.

\section{Energy Resolution Study}
\label{section:EnergyResolution}

\subsection{Data Structure}
\label{subsection:EnergyResolution:DataStructure}

For the energy resolution studies, our detector position was fixed
and data were taken at three
separate accelerator energies, allowing us to measure the energy
resolution for these energies. The three runs that were used for our
results had accelerator energies of $2.537$, $4.446$, and $5.542$ GeV.
Each run had approximately $2 \times 10^{5}$ events.

\begin{figure}
  \centering
  \includegraphics[width=\linewidth]{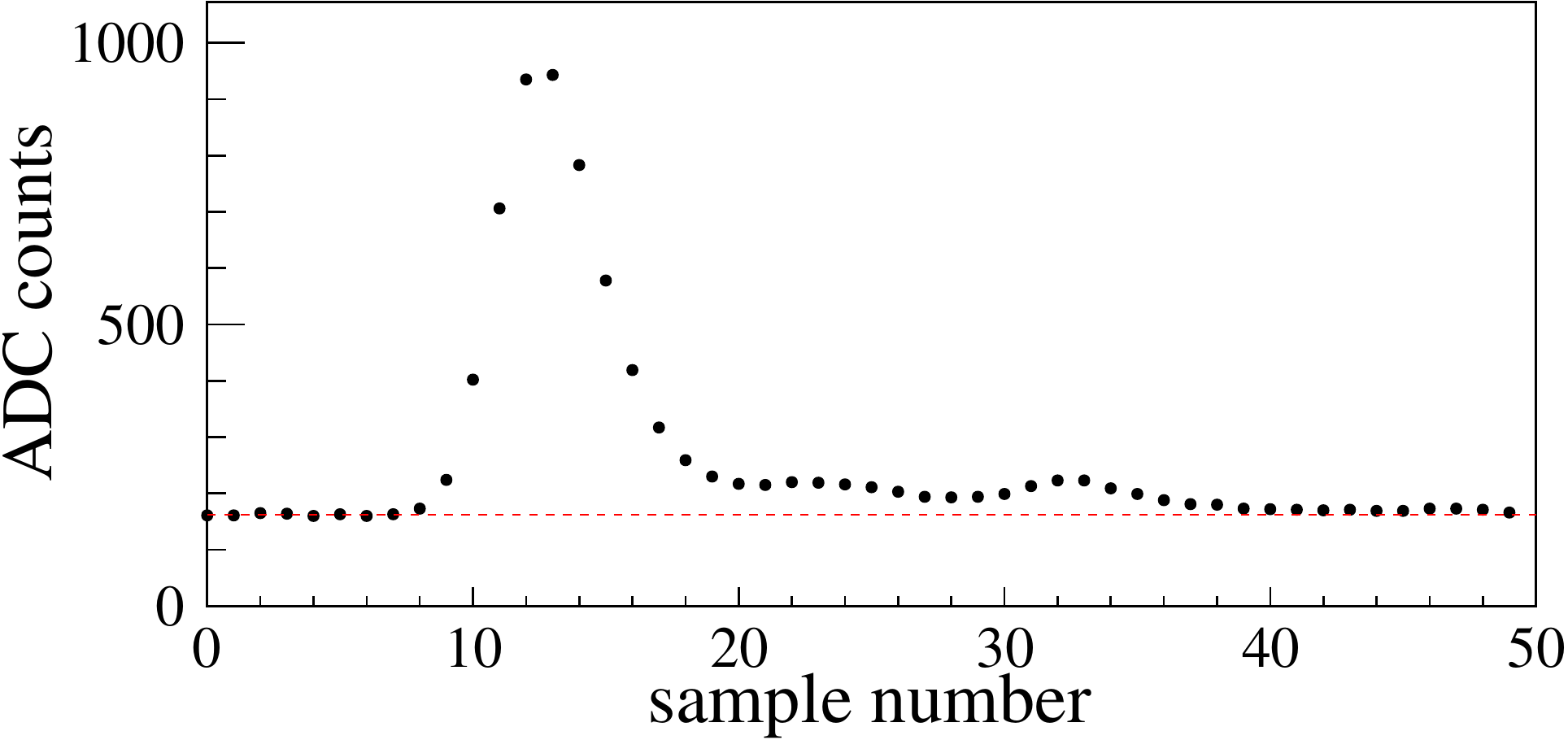}
  \caption{ADC values read out for one channel in one event. The
    pedestal which is given by the average of the first eight values
    is shown as the dashed red line. A clear signal is seen for this
    event.}
  \label{fig:single_eventDisplay}
\end{figure}

Each channel of the fADCs samples the PMT output voltage every $4$ ns,
and $50$ samples were stored for each event.
Figure~\ref{fig:single_eventDisplay} shows the read out samples from a
single channel in one event. However, zero-suppression was implemented
by setting a pre-programmed threshold and only channels that had at
least one signal sample larger than this threshold were written to the
output stream. The threshold applied to each channel was based on the
pedestal, or baseline, which was measured without any input
signals. The threshold was set to $10$ ADC counts above the
pedestal. Off-line studies showed that the resolution worsens with
larger thresholds. After data taking, it was discovered that the
thresholds for modules in the top two rows were inadvertently set much
higher, leading to a significantly worse energy resolution. Therefore,
the sub-sample of events with showers centered on these modules were
not used in the determination of the energy resolution (see
Sec.~\ref{subsection:EnergyResolution:final}).

The module response to energy deposited in the lead glass is obtained
by summing the pedestal-subtracted samples over a range of time. This
range was optimized and was $20$ samples (or 80 ns) starting at sample
$9$. The sum was performed off-line for the beam test, although for
the GlueX experiment the determination of the signal size and time
will be computed in Field-Programmable Gate Arrays (FPGAs) to manage
the size of the output data stream.

The first step in our analysis was to ensure that we had a sample of
events that had one and only one electron signal with the correct
timing. Therefore, a preliminary skim was applied to the data to
extract such events.
For each inside trigger scintillator, we determined the pedestal
by taking the average of the first eight samples recorded.
To exclude events that had an unstable pedestal, we removed
events that had channels with a pedestal RMS greater than
$2$. This eliminated events that were on the
tail of a previous event and also noise fluctuations at the beginning
of the event.

Next, to ensure that the timing of the event was correct,
the recorded NIM signals were examined and a coincidence in the
leading edges was required. Also, events with a prolonged or second
signal were removed.
Finally, we required that only one horizontal and vertical
trigger combination satisfied all of our trigger conditions to ensure
that there were not two separate incoming electrons.

Similar requirements on the pedestal RMS and timing of the signal were made on
the remote paddle based on the distribution of the data.
We verified no signal was present on the veto just before or after the
triggered events.
For all detector modules, if any sample indicated an underflow or
overflow in the ADC counts, that event was removed. The requirements
on the trigger counters removed $10$--$20\%$ of the data, while the
requirements on the remote paddle removed $70$--$80\%$ of our initial
data sample indicating that our detector was frequently triggered by
electrons that didn't pass through narrow angular range that was
spanned by the remote paddle at the exit of the tagger window. 

\subsection{Gain Balancing Procedure}
\label{subsection:EnergyResolution:GainBalancing}

The total signal size for the
event is given by summing
the signals from all modules. For each run,
the distribution of the total signal is
fit with a Gaussian function and linear background. The width
$(\sigma)$ of this distribution divided by the centroid is then the
measured resolution. Below we outline the procedure that was
used to optimize the resolution consistently over all runs.

Prior to the beam test, the gain characteristics for the PMTs for each
module were measured, and the high voltage (HV) settings were
adjusted to equalize the gains. With these initial settings, data were
taken to measure the
initial energy resolution. A software gain balancing technique was then
applied to minimize the event-to-event variance of the sum over all modules
by scaling the signal size from each module by a constant module-dependent gain factor.
 This minimization can be formulated in terms of a
Lagrangian multiplier method, which is described in detail in
Ref.~\cite{Jones_NIM}. In Ref.~\cite{Jones_NIM} the mass squared
of the $\pi^{0}$ and $\eta$ mesons were used as constraints and
measured differences from the meson masses squared were minimized.
We fixed the average total signal size (in arbitrary units), and the standard 
deviation from this average was minimized.  Such a problem is linear with 
respect to the gain factors and can be
solved exactly, so that no iteration of the minimization is necessary.
Next, using the computed gain factors, we adjusted the HV values based on knowledge of the 
gain characteristics of each PMT to better equalize the gains of the PMTs.
Finally, we repeated the gain factor determination, and utilized these 
gain factors in our analysis. 

We  explored the stability of our software gain-balancing procedure by selecting 
events whose sum were within $\pm 3\sigma$ or $\pm 5\sigma$ of the mean and 
determined gain constants for both sets.  This selection reduces the influence of non-Gaussian tails on
the computation of the variance and the gain constants.
Our results are therefore given as the average resolution from these
selections, with an error given by half the difference, which is negligible on
the scale of other systematic uncertainties in the determination of the energy
resolution.

\begin{figure}[h!t!b!p]
  \subfloat{\includegraphics[width=0.90\linewidth]{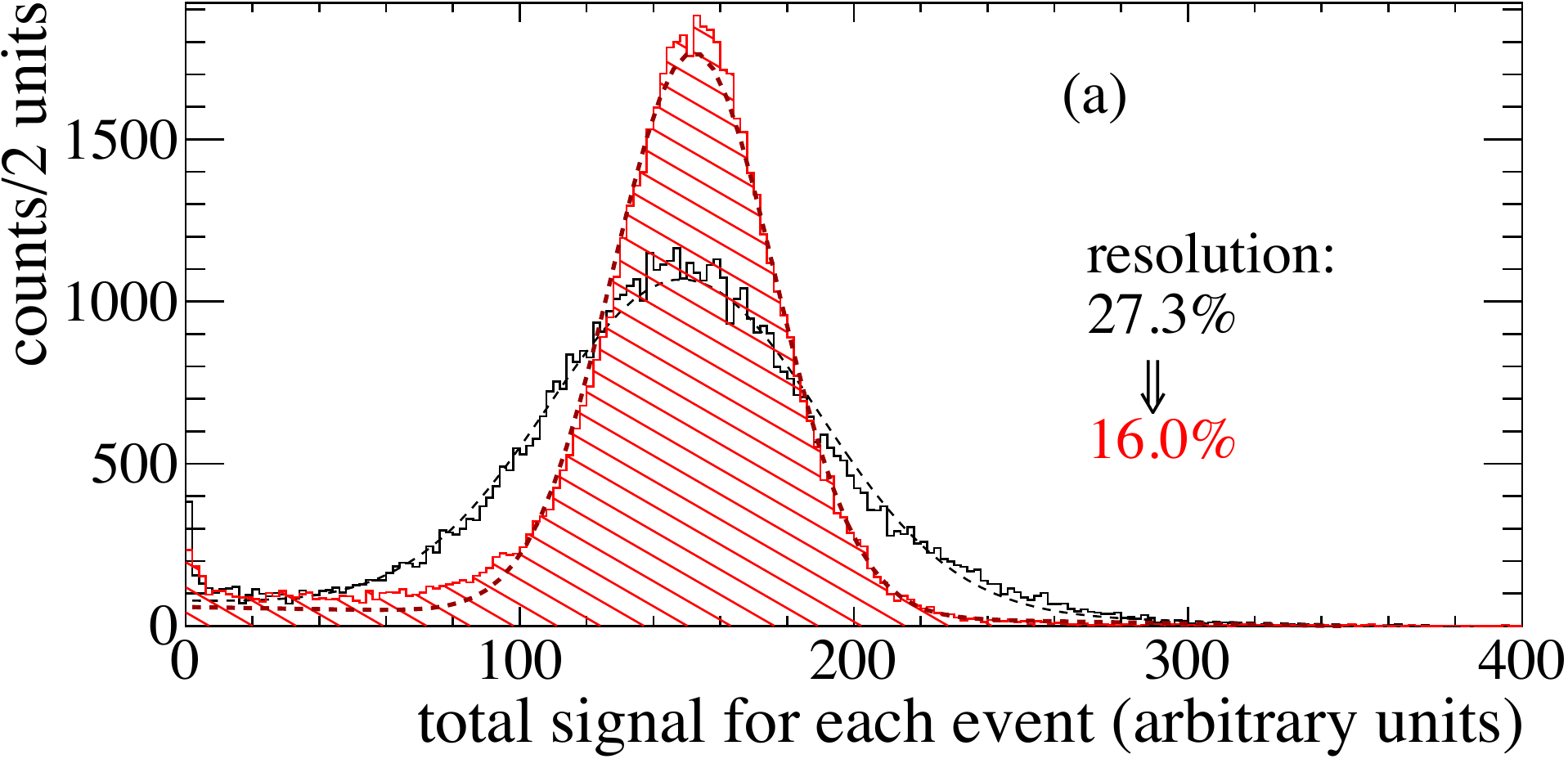}}

  \subfloat{\includegraphics[width=0.90\linewidth]{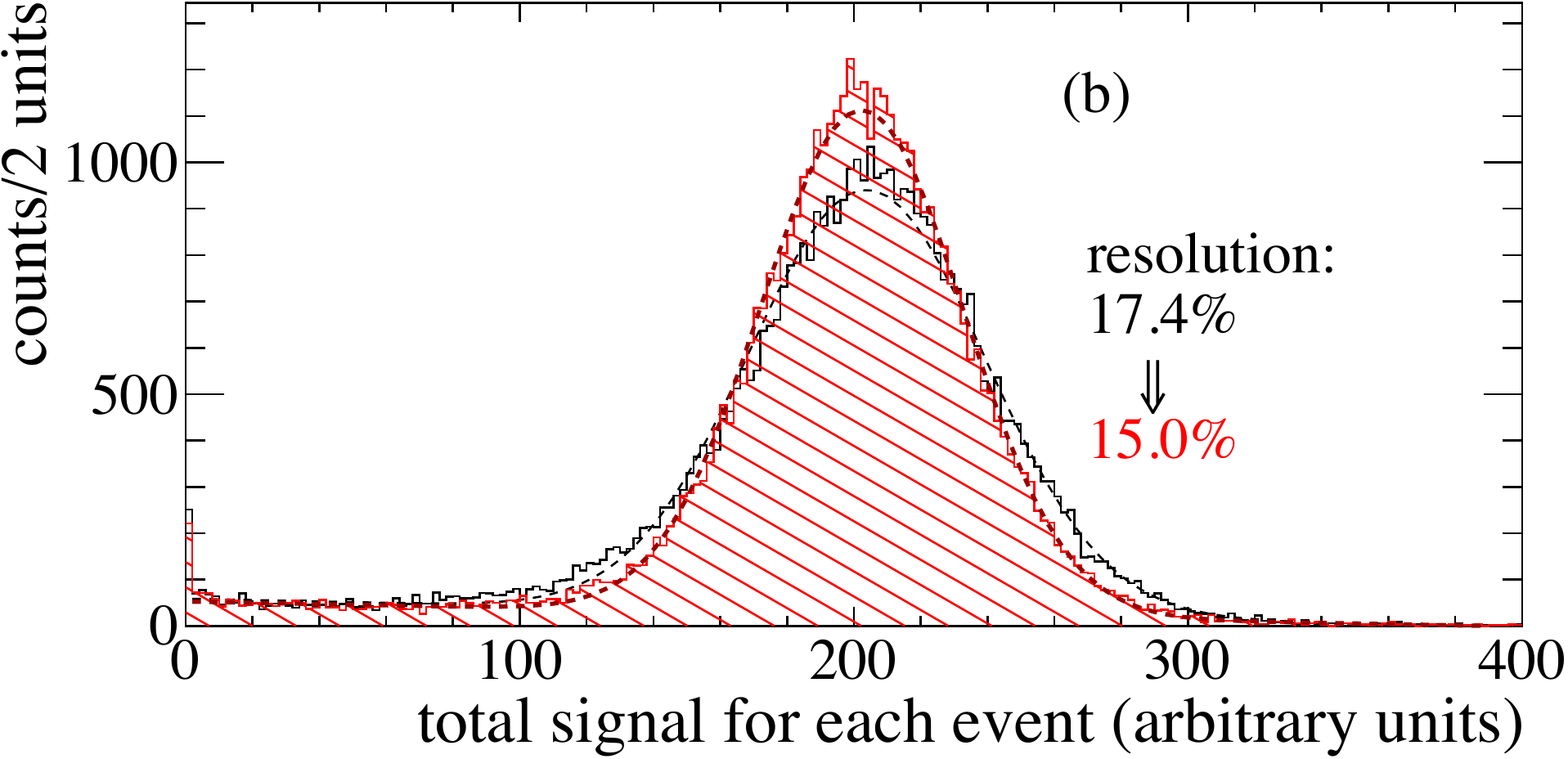}}
  \caption{The distribution of the total signal of all modules before
    (hollow/black online) and after (hatched/red online) software gain balancing.
    Fits with a Gaussian function and linear background are
    shown by dashed lines.
    The two plots are from runs with the
    same electron energy; the top (bottom) plot is before
    (after) PMT HV adjustments.  For the bottom plot, in addition to
    balancing gains, the overall average HV settings for the PMTs were
    raised, which increased the average total signal.}
\label{fig:run474-476_signaldist}
\end{figure}

Figure~\ref{fig:run474-476_signaldist} demonstrates the method of
determining the resolution and shows the distribution of the
total signal from all modules for two given
runs at the same incoming electron
energy. Figure~\ref{fig:run474-476_signaldist}(a) shows a run
in which we used our initial HV settings without the PMT HV adjustments,
and Fig.~\ref{fig:run474-476_signaldist}(b) is for a run when it
was applied. For each case, the black hollow histogram is the distribution before the
software gain balancing procedure is applied, and the gray (red online)
hatched histogram is after.  We see that software gain balancing is essential
to obtaining optimal resolution, and the resolution is best when software
corrections are applied to approximately equal-gain modules to start, {\it i.e.},
the precision that is lost when a low gain module is digitized by the
fADC cannot be recovered by a simple multiplicative constant.  In principle,
one can continue to iterate hardware HV adjustments based on software
gain constants; however, we performed only one such HV adjustment and
utilized the software gain balancing procedure for all subsequently
analyzed data.

\subsection{GEANT Simulations}
\label{subsection:EnergyResolution:GEANT}

Corrections for energy loss and scattering in the material before the
electrons entered the detector were determined using a
GEANT4-based~\cite{GEANT} simulation program. For each simulated data set, nominal
energies for the electrons were chosen to correspond to actual electron energies
used in the data collection. The fractional spread of the actual CEBAF
electron beam is of order $0.01\%$ and is negligible. Electrons were
produced with energies centered
around this value and with a range of $\pm 10\%$ to cover the range of
rays that would scatter into our acceptance. For each event, the
incident angle and position with which the electron exited the vacuum window
were calculated based on the CLAS tagger geometry. The
electron energy distribution was modeled on a realistic bremsstrahlung
distribution.

Using the simulation, we recorded the distribution of total incoming
energy into our detector.  The same trigger conditions as
used in the analysis of the data were imposed on the simulated events.
We fit the incoming energy distribution for our final
selection of events with a Gaussian function to determine the mean and
width.
We concluded that the mean
incoming energy into our detector was $12$--$14$ MeV lower than the
nominal energy with a spread of $4$--$6$
MeV. Table~\ref{tab:simulation_results} summarizes the results of our
simulation for all energies. With the assumption that the intrinsic
detector resolutions and the spread of energies add in quadrature, we
can subtract this contribution to obtain the intrinsic detector
resolution.

\begin{table}[h!t!b!p]
  \centering
  \caption{\label{tab:simulation_results}
    Predicted energy and spread of electrons incident on the detector
    during three run periods.
    $E_{0}$ is the nominal incoming energy, while
    $E$ and $\sigma_{\text{loss}}$ are the
    center and width of the distribution obtained from the
    simulation, respectively.
  }
  \begin{tabular}{rccc}\hline \hline
    $E_{0}$ (MeV) & $E$ (MeV) & $\sigma_{\text{loss}}$ (MeV) & $\sigma_{\text{loss}}/E$ (\%) \\ \hline
    $126.85$ & $114.19 \pm 0.03$ & $4.54 \pm 0.02$ & 3.98 \\
    $222.30$ & $209.29 \pm 0.05$ & $5.15 \pm 0.03$ & 2.46 \\
    $277.10$ & $263.70 \pm 0.04$ & $5.70 \pm 0.03$ & 2.16 \\ \hline \hline
  \end{tabular}
\end{table}

We incorporate these simulation results into our analysis in the following
way. The spread in measured energies, $\sigma_{\text{meas}}$, is the
quadratic sum of the intrinsic detector resolution,
$\sigma_{\text{det}}$, and the spread due to different incoming energies, which we
approximate with our simulation results, $\sigma_{\text{loss}}/E$.  With the
assumption that the intrinsic detector resolution and the spread of
energies adds in quadrature, we can
remove the spread of energies to obtain the intrinsic detector
resolution.

\subsection{Energy Resolution}
\label{subsection:EnergyResolution:final}

\begin{figure}[t!b!p]
  \centering
  \includegraphics[width=0.95\linewidth]{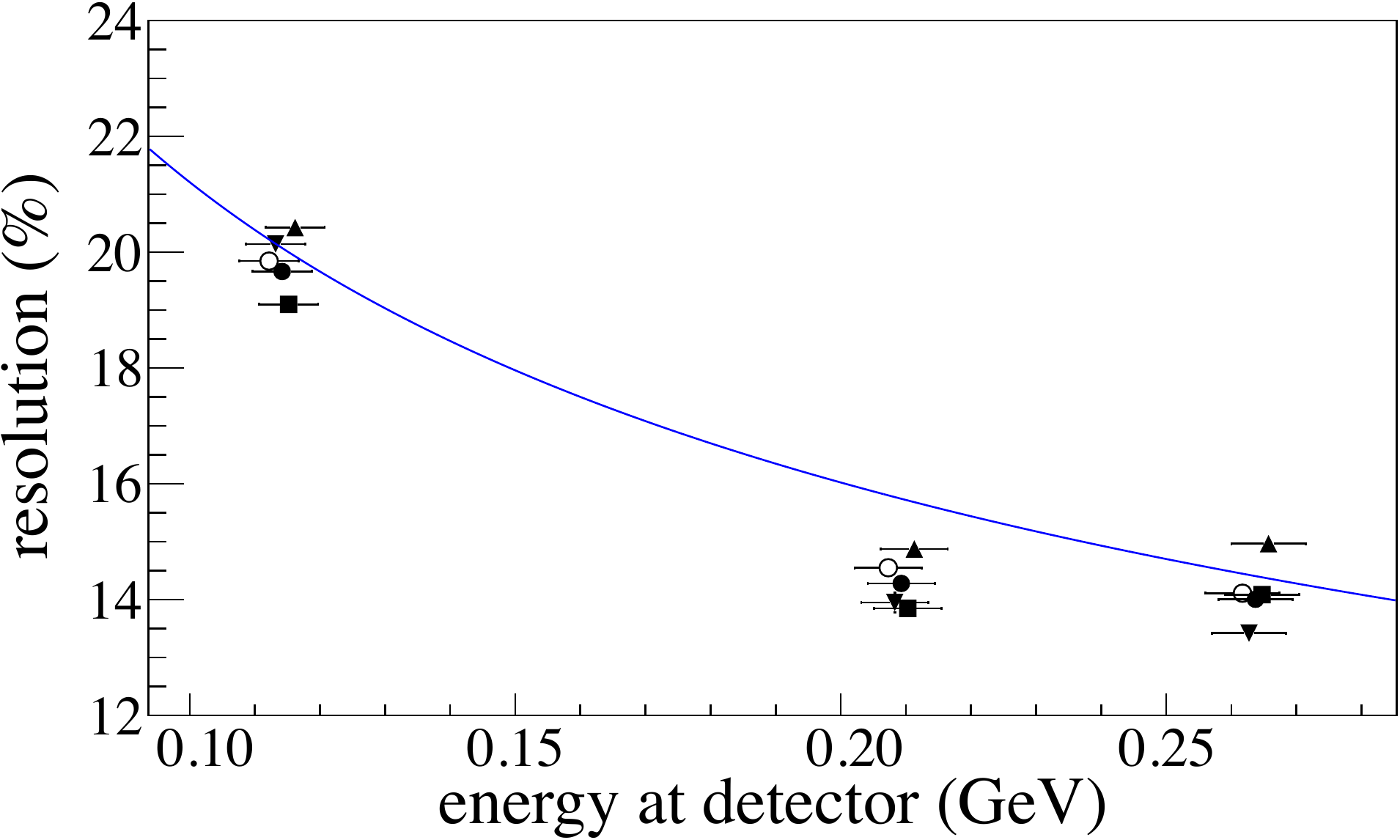}
  \caption{The energy resolution results against
    incoming energy. The different markers represent different
    trigger combination selections, and are slightly
    horizontally shifted for visual clarity. The solid blue curve is
    the expected resolution for GlueX.}
  \label{fig:finalResolution}
\end{figure}

Figure~\ref{fig:finalResolution} shows the energy resolution for
five different trigger combinations. For all runs, the statistical uncertainties from the fits to
determine $\sigma_\mathrm{meas}$ were negligibly small compared to
the systematic variations observed by changing the analysis method.
The dominant source of systematic uncertainty is variation in the performance of the
individual modules.  The nine different trigger possibilities, 
3 horizontal $\times$ 3 vertical, could be analyzed independently and each
trigger combination populates the individual modules differently.  Therefore,
the energy resolution varies with the location of the incident electron.  
Three trigger combinations were excluded due to the incorrect hardware 
threshold settings and a
fourth combination was also excluded because it consistently gave poor
resolution, although a clear reason was not identified. Each marker
represents a different trigger combination, and have been shifted
horizontally slightly for visual clarity. The measured energy
resolution for the three electron energies of  $114.2
\pm 4.5$~MeV, $209.3 \pm 5.2$~MeV, and $263.7 \pm 5.7$~MeV, as given
by the average and standard deviation of the five trigger
combinations, are $19.8 \pm 0.5 \%$, $14.3\pm 0.4\%$, and $14.1 \pm
0.5\%$, respectively. 

We expect that the dominant contribution to the resolution in this energy regime is statistical
fluctuations in the number of photoelectrons produced in each PMT
for each electron shower.  Based on simulations of the light
collection optics that were carried out in the design phase, the expected resolution
of the GlueX FCAL can be parametrized as
\begin{align}
  \frac{\sigma}{E} (\%)&= \frac{5.6}{\sqrt{E (\text{GeV})}} +
  3.5. \label{eq:RadPhi}
\end{align}
This is shown as the solid blue curve in
Fig.~\ref{fig:finalResolution}.  Our energy resolution
results are consistent with the design goal in this energy regime.

\section{Timing Resolution Study}
\label{section:TimingResolution}

The second important performance metric of the calorimeter is the
resolution of the absolute arrival time of the PMT pulse that is digitized by the fADCs.
The fADCs for the FCAL sample the PMT pulses only every $4$ ns; however,
it is possible to utilize several samples and knowledge about the pulse shape to
determine a pulse time with resolution much better than 4~ns. In a previous
article~\cite{IU-NIM}, the timing resolution of the fADCs was tested
using a pulsed LED source, and a timing resolution of
less than $1$ ns was determined for pulse heights of $100$~mV and
higher. In the following, we apply the same
methods as in Ref.~\cite{IU-NIM} for the purpose of confirming that
we can obtain adequate timing resolution using actual electromagnetic
showers. For this analysis, we utilize a single run with $507{,}828$ events that was taken
at an incoming electron energy of approximately $1275$ MeV.
The timing resolution depends strongly on the PMT signal amplitude,
and the runs at this energy allowed us to measure
the timing resolution across the largest range of sample pulse heights.

\subsection{Timing resolution method}
An example of the digitized PMT signal by the fADCs was shown in
Fig.~\ref{fig:single_eventDisplay}. We use the time at
which the pulse reaches half the recorded maximum to specify the
signal arrival time. This time, which we call $t_{0}$, is chosen
because it is in the region of maximum slope.

The timing resolution was determined using a linear
interpolation method, which assumes that we can approximate the
signal shape between adjacent samples of the fADC linearly, a valid
approximation on the leading edge of the pulse.  We
determined $t_{0}$ by interpolating the sample before and the sample
after the halfway point of the maximum sample.
Generally, the incoming electron in the beam test
illuminated several blocks simultaneously, allowing us to compute the 
difference in $t_{0}$ for several different modules.  We define the
timing difference between two modules as
$\Delta t_{0,ij} = t_{0,i} - t_{0,j}$,
where $t_{0,i}$ is the signal arrival time $t_{0}$ for the $i$th
module.

\begin{figure}[h!t!]
  \centering
  \includegraphics[width=0.90\linewidth]{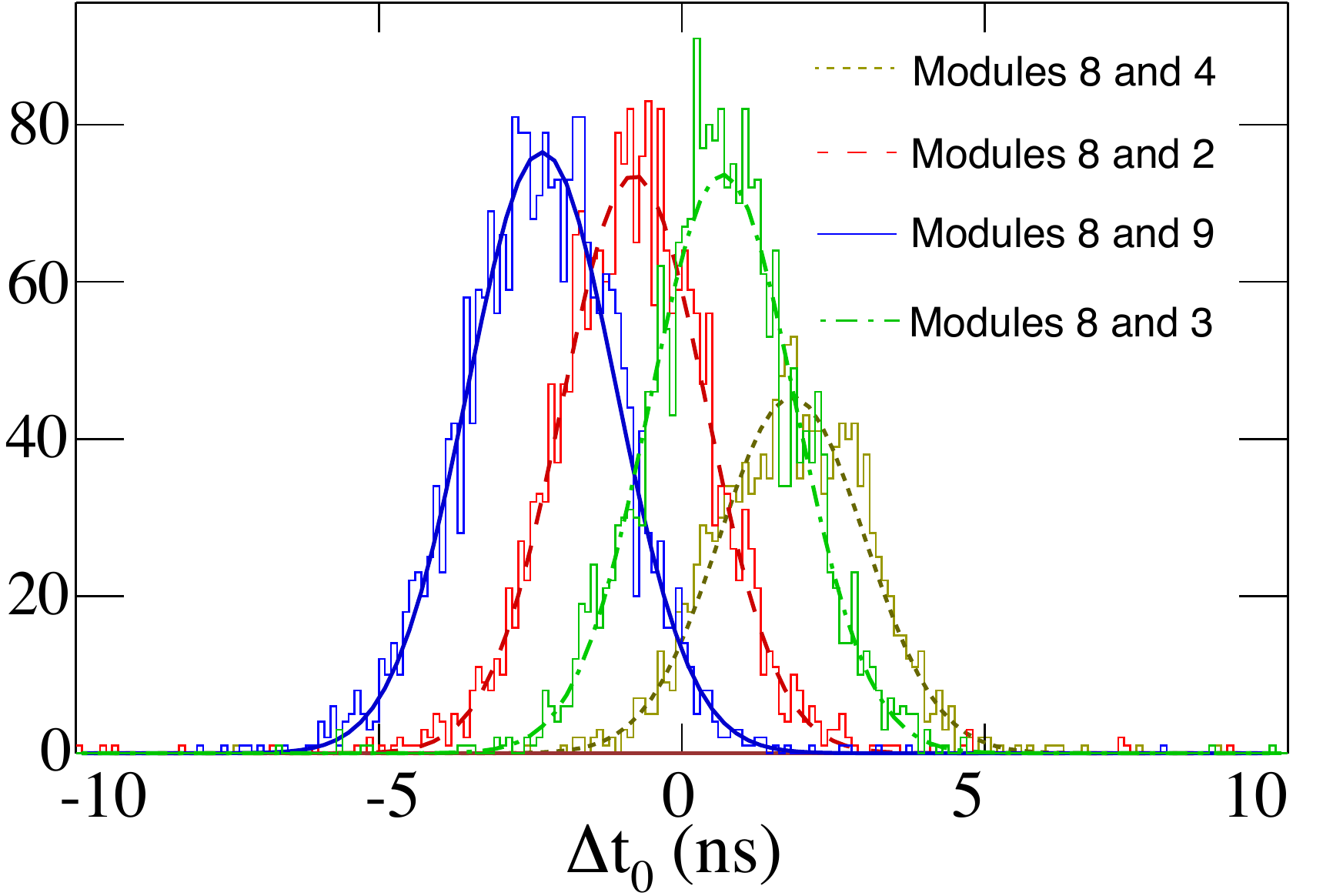}
  \caption{Distribution of $\Delta t_{0,ij}$ for a single module and
    four of its adjacent modules when all modules had $1000 < S_{p} <
    2000$ ADC counts, together with
    Gaussian fit curves.}
\label{fig:timing_dist}
\end{figure}

In principle we can compute the timing difference between any two
modules, but since we are interested in determining the resolution
for two modules with comparable signal amplitudes, we restricted
ourselves to measurements of differences between adjacent modules,
modules that are either side by side or diagonally
aligned with the corners touching. Figure~\ref{fig:timing_dist} shows
the distribution of the timing difference $\Delta t_{0,ij}$ for a
single module and four of its adjacent modules. Previous studies
showed that the timing resolution depends strongly on the maximum
pulse amplitude $S_{p}$, so the
data were binned by requiring that for a given event, $S_{p}$ for
both modules were within the same range. For example in
Fig.~\ref{fig:timing_dist}, the range of $S_{p}$ was restricted to
be between $1000$ and $2000$ ADC counts for the two modules.

\subsection{Determination of timing resolution}

\begin{figure}[h!t!]
  \centering
  \includegraphics[width=0.95\linewidth]{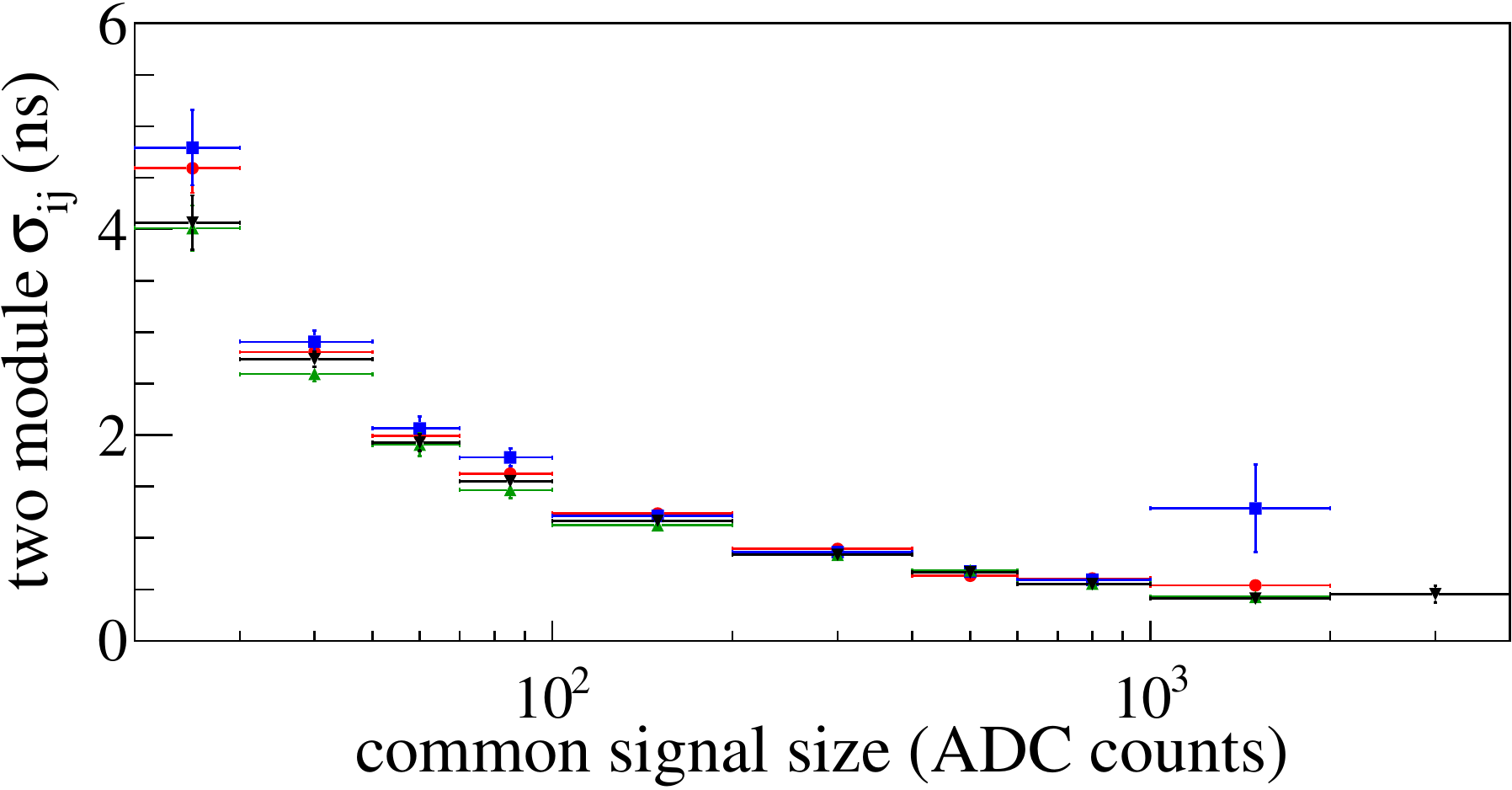}
  \caption{$\sigma_{ij}$ between an
    example module and four of its adjacent modules,
    determined by fits such as those shown
    in Fig.~\ref{fig:timing_dist}. The
    four different colors and symbols represent different combinations
    of modules. Note the log scale on the $x$-axis.
    }
\label{fig:timing_diff_sigma}
\end{figure}

For each range of $S_{p}$, the distribution of $\Delta t_{0,ij}$ was
determined for all adjacent module combinations. Fits to each
distribution were done with Gaussian functions, such as those shown
in Fig.~\ref{fig:timing_dist}, to characterize each $\Delta
t_{0,ij}$, which yield the resolution in timing difference
$\sigma_{ij}$ with statistical error $\delta
\sigma_{ij}$. Fig.~\ref{fig:timing_diff_sigma} shows how the
$\sigma_{ij}$ between a given module and four of its adjacent modules
changes with $S_{p}$. Clearly the
timing resolution improves with larger pulse amplitude.

The value $\sigma_{ij}$ is the
resolution on the difference in absolute time between two modules and is therefore
worse than the time resolution for a single module. We assumed that the resolution of
each single module
$\sigma_{i}$ contributes in quadrature to the observed timing
difference resolution so that $\sigma_{ij}^{2} = \sigma_{i}^{2} +
\sigma_{j}^{2}$, and determined each module's
resolution from all of the different combinations measured. To
do this, we divided our modules into $2 \times 2$ clusters where all
modules within the cluster were adjacent to each other. Within a
single cluster there are six possible separate measurements
of timing differences. If all six combinations each had sufficient
statistics to yield a fit result, we did a fit to minimize
\begin{align}
  \chi^{2} &= \sum_{i,j>i} \left( \frac{\sigma_{ij} -
    \sqrt{\sigma_{i}^{2} + \sigma_{j}^{2}}}{\delta \sigma_{ij}}
  \right)^{2},
\end{align}
where the fit variables $\sigma_{i}$ gave the intrinsic timing
resolution of module $i$. This allowed a determination of the
resolution of each module within that cluster.

\begin{figure}[h!t!b!p!]
  \centering
  \includegraphics[width=\linewidth]{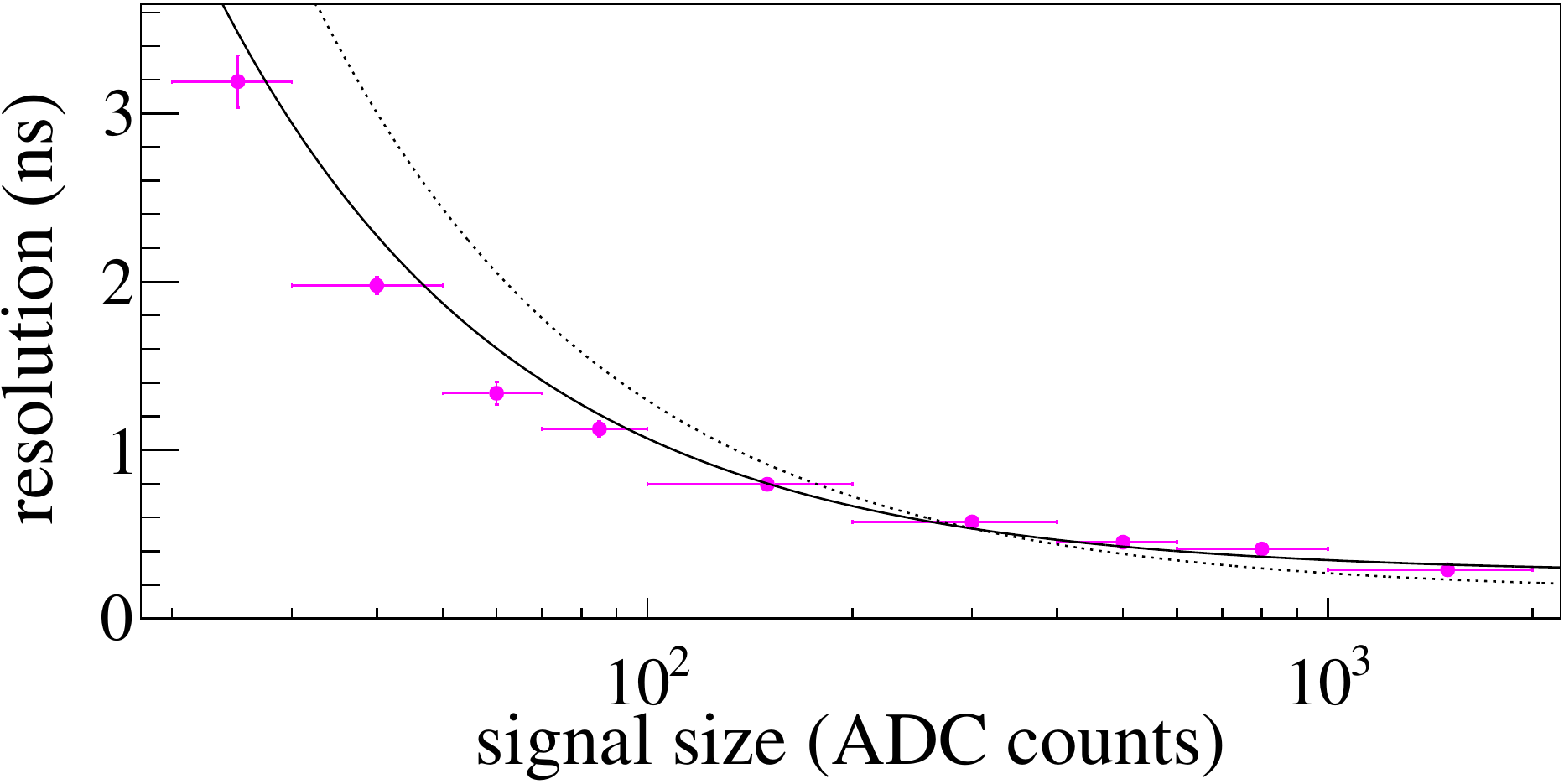}
  \caption{Final timing resolution for one module.
    The solid magenta points are
    the weighted averages and standard deviations of up to four
    separate clusters that 
    determined the resolution for this module. The
    solid curve shows the fit result, while the dotted curve is from
    Ref.~\cite{IU-NIM}.}
\label{fig:timing_sigma}
\end{figure}

Figure~\ref{fig:timing_sigma} shows the resolution of a single
module. Because each module can be contained in up to four separate
clusters, the timing resolution in a given bin of $S_{p}$ can be
determined  up to four times. When there were
multiple measurements coming from different clusters at a given range
of $S_{p}$, the different measurements were merged by a weighted
average, with the standard deviation taken as the error. We
parametrized the results with fits of the form
\begin{align}
  \sigma(S_{p}) &= \frac{a}{S_{p}} + b, \label{eq:oneoverx}
\end{align}
where $a$ ($\text{ns} \cdot \text{ADC counts}$) and $b$ ($\text{ns}$) are fit
parameters. Fits were also tried with other
forms, such as $a/\sqrt{S_{p}} + b$ and $a/S_{p} \oplus b$, where in the
latter form the two terms were added in quadrature, but neither of these
fit the data better than Eq.~\eqref{eq:oneoverx}.

By fitting the timing resolution as a function of $S_{p}$, we
determined the parameters $a$ and $b$ for each module.
Table~\ref{tab:timing_results} shows the results of the parameters $a$
and $b$ for the six modules we used. By taking the average and
standard deviation, our final results were $a = 72.9 \pm 5.3$
($\text{ns} \cdot \text{ADC counts}$), $b = 0.29 \pm 0.02$ ($\text{ns}$).
We can convert the ADC counts into pulse heights in mV:
the fADCs used $12$ bits to digitize the $0.5$~V fADC full scale.
With this conversion, the fit
parameter $a$ becomes $a = 8.90 \pm 0.65$ $\text{ns} \cdot
\text{mV}$.  We note that, on average, the gain of the modules is such
that a signal pulse amplitude of 3 ADC counts corresponds to 1~MeV
of deposited energy in a block. 

The results from Ref.~\cite{IU-NIM} gave the fit parameters $a = 114
\pm 46$ ($\text{ns} \cdot \text{ADC counts}$), $b = 0.155 \pm 0.077$
($\text{ns}$). The statistical parameter can be converted to pulse
height with the $1.45$~V fADC full scale used, and gives $a = 40.4 \pm
16.3$ $\text{ns} \cdot \text{mV}$. For pulse heights of $100$~mV and
$500$~mV, the timing resolutions are then $0.57 \pm 0.18$~ns, and
$0.24 \pm 0.08$~ns, respectively. In the current measurements the
actual pulse heights were limited to less than approximately $250$ mV,
but extrapolating our fit results gives timing resolutions of
$0.38 \pm 0.03$~ns and $0.30 \pm 0.02$~ns for $100$~mV and $500$~mV pulses,
respectively. Our results show that the timing resolution exceeded
expectations at smaller signal amplitudes where previous results do
not exist.

\begin{table}[h!t!bp]
  \centering
  \caption{\label{tab:timing_results} Measurements of timing
    resolution parameters $a$ and $b$ from Eq.~\eqref{eq:oneoverx}.
    The final row shows the average values and standard deviations of
    the six modules used.}
  \begin{tabular}{ccc} \hline \hline
    Module number &  $a$ ($\text{ns} \cdot \text{ADC counts}$) & $b$ (ns) \\ \hline
    7  & $77.9 \pm 8.5$ & $0.27 \pm 0.02$\\
    8  & $65.6 \pm 6.7$	& $0.32 \pm 0.02$\\
    9  & $78.9 \pm 8.1$ & $0.31 \pm 0.02$\\
    12  & $68.1 \pm 6.8$ & $0.31 \pm 0.02$\\
    13  & $72.5 \pm 7.7$ & $0.28 \pm 0.02$\\
    14  & $74.6 \pm 7.8$ & $0.28 \pm 0.02$ \\
    \textbf{Average} & $\mathbf{72.9 \pm 5.3}$ & $\mathbf{0.29 \pm 0.02}$ \\ \hline \hline
  \end{tabular}
\end{table}

In the GlueX experiment, a single photon can cause showers in multiple
blocks, which allow independent timing measurements from each block.
Combining these measurements will further enhance the timing resolution of
photon showers, thereby enhancing our ability to choose which beam
bunch produced the signal photon of interest and reject background
electromagnetic showers.

\section{Conclusions}

Results of a beam test using a small version of the FCAL of the GlueX
experiment are given. The energy resolution is between $14\%$ and
$20\%$ at energies of $260$~MeV down to $110$~MeV.
This represents the first test of the production calorimeter hardware
and electronics with actual electromagnetic showers.
The energy resolution is consistent with the design goals for GlueX at these low energies.
Precision timing measurements of the FCAL signals is a
feature that is possible due to the $250$~MHz fADCs
to be used in GlueX. The timing resolution is found to exceed the expectations of previous
measurements at low PMT pulse amplitudes, with timing resolutions of
$0.38$~ns and better achievable from a single module with signals
larger than $100$~mV. A weighted average of timing measurements
from different modules that are illuminated by a single shower will
allow enhanced precision of shower time allowing a determination of the 
beam bunch and a rejection of out-of-time electromagnetic background.

\section*{Acknowledgments}

We would like to thank the CLAS Collaboration
and the members of the HD-ICE experiment, for their
hospitality, time and effort, which made our beam test possible.
We would especially like to thank Eugene Pasyuk and Sergey Boyarinov
for their help.
% IU contract
This work was supported by the Department of Energy under contract
DE-FG02-05ER41374.
% JLab DOE contract
Jefferson Science Associates, LLC operated
Thomas Jefferson National Accelerator Facility for the United
States Department of Energy under contract
DE-AC05-06OR23177.


\begin{thebibliography}{99}
\bibitem{GlueX-URL} \url{https://www.jlab.org/Hall-D}
\bibitem{CLAS-URL} \url{https://www.jlab.org/Hall-B}
\bibitem{Brunner} A. Brunner et al., A Cockcroft-Walton base for the FEU84-3 photomultiplier
                        tube, Nuclear Instruments and Methods in
                        Physics Research A 414 (1998) 466. doi:10.1016/S0168-9002(98)00651-2
\bibitem{Lytkarino} \url{http://lzos.ru/}
\bibitem{fADC} H. Dong et al., Integrated tests of a high speed VXS
  switch card and 250 MSPS flash ADCs, Nuclear Science Symposium Conference
  Record, 2007. NSS '07. IEEE, Vol. 1, 2007, pp. 831-833. doi:10.1109/NSSMIC.2007.4436457
\bibitem{Brabson} B.B. Brabson et al., A Study of two prototype lead glass electromagnetic
                        calorimeters, Nuclear Instruments and Methods
                        in Physics Research A 332 (1993)
                        419. doi:10.1016/0168-9002(93)90299-W
\bibitem{Crittenden} R.R. Crittenden et al., A 3000 element lead-glass
  electromagnetic calorimeter, Nuclear Instruments and Methods in
  Physics Research A 387 (1997)
  377. doi:10.1016/S0168-9002(97)00101-0
\bibitem{Jones_NIM} R.T. Jones et al., A bootstrap method for gain calibration and resolution
                        determination of a lead-glass calorimeter,
                        Nuclear Instruments and Methods in Physics
                        Research A 566 (2006) 366. doi:10.1016/j.nima.2006.07.061
\bibitem{Jones_NIM2} R.T. Jones et al., Performance of the RADPHI detector and trigger in a high
                        rate tagged photon beam,
                        Nuclear Instruments and Methods in Physics
                        Research A 570 (2007) 384. doi:10.1016/j.nima.2006.09.039
\bibitem{tagger}  D.I. Sober et al., The bremsstrahlung tagged photon
  beam in Hall B at JLab, Nuclear Instruments and Methods in Physics
  Research A 440 (2000) 263. doi:10.1016/S0168-9002(99)00784-6
\bibitem{vacuum_window} S.K. Matthews et al., A composite thin vacuum window for the CLAS photon
                        tagger at Jefferson lab,
                        Nuclear Instruments and Methods in Physics
                        Research A 421 (1999) 23. doi:10.1016/S0168-9002(98)00910-3
\bibitem{GEANT} S.Agostinelli et al., GEANT4: A Simulation toolkit,
  Nuclear Instruments and Methods in Physics Research A 506 (2003) 250. doi:10.1016/S0168-9002(03)01368-8
\bibitem{IU-NIM} J.V. Bennett et al., Precision timing measurement of phototube pulses using a
                        flash analog-to-digital converter,
                        Nuclear Instruments and Methods in Physics
                        Research A 622 (2010) 225. doi:10.1016/j.nima.2010.06.216
\end{thebibliography}
\end{document}